\documentclass{aa}
\usepackage{graphicx}
\usepackage{rotating}
\begin{document}

\newcommand{\eff}{\mbox{Effelsberg}}
\newcommand{\gb}{\mbox{Green Bank}}
\newcommand{\nan}{\mbox{Nan\c{c}ay}}
\newcommand{\as}[2]{$#1''\,\hspace{-1.7mm}.\hspace{.1mm}#2$}
\newcommand{\am}[2]{$#1'\,\hspace{-1.7mm}.\hspace{.0mm}#2$}
\newcommand{\dgr}{\mbox{$^\circ$}}
\newcommand{\E}[1]{\mbox{${}\,10^{#1}{}$}}
\newcommand{\ea}{{\it et al.}}
\newcommand{\grd}[2]{\mbox{#1\fdg #2}}
\newcommand{\gsim}{\stackrel{>}{_{\sim}}}
\newcommand{\lsim}{\stackrel{<}{_{\sim}}}
\newcommand{\Ha}{\mbox{H$\alpha$}}
\newcommand{\HI}{\mbox{H\,{\sc i}}}
\newcommand{\HIbf}{\mbox{H\hspace{0.155 em}{\footnotesize \bf I}}}
\newcommand{\HIit}{\mbox{H\hspace{0.155 em}{\footnotesize \it I}}}
\newcommand{\HIsl}{\mbox{H\hspace{0.155 em}{\footnotesize \sl I}}}
\newcommand{\HIss}{\mbox{H\,{\sc i}}}
\newcommand{\HII}{\mbox{H\,{\sc ii}}}
\newcommand{\Jykms}{\mbox{\rm Jy km\,s$^{-1}$}}
\newcommand{\kms}{\mbox{\rm km\,s$^{-1}$}}
\newcommand{\kmsMpc}{\mbox{\rm km\,s$^{-1}$\,Mpc$^{-1}$}}
\newcommand{\LB}{\mbox{$L_{B}$}}
\newcommand{\LBnul}{\mbox{$L_{B}^0$}}
\newcommand{\LBsun}{\mbox{$L_{\odot,B}$}}
\newcommand{\Lsun}{\mbox{$L_{\odot}$}}
\newcommand{\LsunMsun}{\mbox{$L_{\odot}$/${M}_{\odot}$}}
\newcommand{\LFIR}{\mbox{$L_{FIR}$}}
\newcommand{\LFIRLB}{\mbox{$L_{FIR}$/$L_{B}$}}
\newcommand{\LFIRLBnul}{\mbox{$L_{FIR}$/$L_{B}^0$}}
\newcommand{\LFIRLsun}{\mbox{$L_{FIR}$/$L_{\odot}$}}
\newcommand{\MHI}{\mbox{${M}_{HI}$}}
\newcommand{\MHILB}{\mbox{${M}_{HI}$/$L_{B}$}}
\newcommand{\MHILBnul}{\mbox{${M}_{HI}$/$L_{B}^0$}}
\newcommand{\Msun}{\mbox{${M}_\odot$}}
\newcommand{\MsunLsun}{\mbox{${M}_{\odot}$/$L_{\odot}$}}
\newcommand{\MsunLBsun}{\mbox{${M}_{\odot}$/$L_{\odot,B}$}}
\newcommand{\MT}{\mbox{${M}_{\rm T}$}}
\newcommand{\MTLBnul}{\mbox{${M}_{T}$/$L_{B}^0$}}
\newcommand{\MTLBsun}{\mbox{${M}_{T}$/$L_{\odot,B}$}}
\newcommand{
\mi}{\mbox{$\mu$m}}
\newcommand{\NH}{\mbox{N$_{HI}$}}
\newcommand{\OIII}{\mbox{[O\,{\sc iii}]}}
\newcommand{\s}{\mbox{$\sigma$}}
\newcommand{\Tb}{\mbox{$T_{b}$}}
\newcommand{\tis}[2]{$#1^{s}\,\hspace{-1.7mm}.\hspace{.1mm}#2$}
\newcommand{\vhel}{\mbox{$V_{hel}$}}
\newcommand{\vrot}{\mbox{$V_{rot}$}}
\def\la{\mathrel{\hbox{\rlap{\hbox{\lower4pt\hbox{$\sim$}}}\hbox{$<$}}}}
\def\ga{\mathrel{\hbox{\rlap{\hbox{\lower4pt\hbox{$\sim$}}}\hbox{$>$}}}}

  \title{Non-confirmation of reported H{\Large \bf I} clouds without optical counterparts in the Hercules Cluster}

    \author{W. van Driel \inst{1},
              K. O'Neil \inst{2},
            V. Cayatte \inst{1},
            P.-A. Duc \inst{3},
           J.M. Dickey \inst{4},
            C. Balkowski \inst{1},
             H. Hern\'andez \inst{2},
            J. Iglesias-P\'{a}ramo \inst{5},
            P. Papaderos \inst{6},
            J.M. V\'{\i}lchez \inst{7}
             \and
            T.X. Thuan \inst{8}
            }

\offprints{W. van Driel}

  \institute{Observatoire de Paris, Section de Meudon, GEPI, CNRS UMR 8111, 
              5 place Jules Janssen, F-92195 Meudon, France \\
             \email{wim.vandriel@obspm.fr, veronique.cayatte@obspm.fr, chantal.balkowski@obspm.fr }
        \and
              Arecibo Observatory, HC3 Box 53995, Arecibo, Puerto Rico 00612, U.S.A. \\
             \email{koneil@naic.edu, hhernand@naic.edu}
        \and
             CNRS URA 2052 and CEA,DSM,DAPNIA, Service d'astrophysique,
            Centre d'Etudes de Saclay, F-91911 Gif sur Yvette cedex, France \\
             \email{paduc@cea.fr}
         \and
             Department of Astronomy, University of Minnesota, Minneapolis, U.S.A. \\
             \email{john@astro.umn.edu}
         \and
            Laboratoire d'Astrophysique de Marseille, Traverse du Siphon - 
            Les Trois Lucs, F-13376 Marseille, France \\
            \email{jorge.iglesias@oamp.fr}
         \and
           Universit\"{a}ts-Sternwarte, Geismarlandstrasse 11, 37083 G\"{o}ttingen, Germany\\
            \email{papade@uni-sw.gwdg.de}
         \and
            Instituto de Astrof\'{\i}sica de Andaluc\'{\i}a (CSIC), Granada, Spain\\
            \email{jvm@iaa.es}
         \and
             Astronomy Department, University of Virginia, Charlottesville, VA 22903, U.S.A. \\
             \email{txt@astro.virginia.edu}
             }


   \date{\it Received 30/8/2002; accepted 8/11/2001}

  \abstract{{\rm
21 cm \HI\ line observations were made with the Arecibo Gregorian telescope of 9 \HI\ clouds
in the Hercules Cluster which were reported as tenative detections in a VLA \HI\ study of the cluster 
(Dickey 1997) and for which our deep CCD imaging failed to find any optical counterparts.
No sensitive observations could be made of one of these (sw-174) due to the presence of a close-by
strong continuum source.  The other 8 tentative \HI\ detections were not confirmed by the Arecibo
\HI\ measurements. The CCD images did reveal faint, low surface brightness counterparts near the centres
of two other VLA \HI\ sources invisible on the Palomar Sky Survey, sw-103 and sw-194.
}
     }

 \authorrunning{W. van Driel et al. }
 \titlerunning{Non-confirmation of \HI\ clouds without optical counterparts in the Hercules Cluster}

 \maketitle

\section{Introduction}
Intergalactic \HI\ clouds have been found to be extremely rare, although they were 
searched for rather intensively in the 21 cm line over the years, in the field as well 
as in environments with higher galaxy density, using both pointed and ``blind'' surveys
(see Briggs 1990 for a summary of the surveys made until 1989 and, e.g., Spitzak \& Schneider 1998; 
Kraan-Korteweg et al. 1999; Zwaan 2001). Their space density was quantified by Briggs (1990) as
less than 1\% of that of galaxies in the \HI\ mass range of about 10$^8$ to 10$^10$ \Msun, based
on data available at the time, while Zwaan (2001) estimated that in groups of galaxies their 
total \HI\ mass is less than 10\% of the total \HI\ mass of the group.

Furthermore, almost all of the claimed cases turned out to have
optical counterparts, e.g., the `isolated' cloud \HI\ 1225+01 discovered at Arecibo by Giovanelli \&
Haynes (1989), which was later found to be associated with a blue dwarf galaxy
(Djorgovski 1991; Salzer et al. 1992). On the other hand, the large intergalactic
cloud discovered at Arecibo by Schneider et al. (1983) in the Leo group has a ring
shape typical of tidal features and may well have been tidally stripped from one of
the galaxies in its vicinity. An example of such a tidal cloud without an optical
counterpart can also be found in the Local Group:  the Magellanic Stream. A new \HI\
cloud without an optical counterpart found by Kilborn et al. (2000) could have been
ejected from the interacting Magellanic Cloud - Galaxy system, and clouds once thought to
be in the Sculptor group (Mathewson et al. 1975) were later shown to be part of the
Magellanic Stream (Haynes \& Roberts 1979).
Although in the Hydra cluster we confirmed (Duc et al. 1999) the VLA \HI\ source 
H1032-2819, for which we did not find an optical counterpart in CCD images,
it might be the result of an interaction between two cluster galaxies.
In a VLA survey of the Coma cluster (Bravo-Alfaro et al. 2000, 2001) two \HI\ clouds 
without optical counterparts on the DSS were found with an \MHI\ of 1 and 4 10$^8$ \Msun,
respectively.

In a 21 cm \HI\ line survey of the Hercules Cluster carried out with the VLA by Dickey
(1997, hereafter D97) the tentative detection was reported of 12 \HI\ clouds without
optical counterparts on digitised Palomar Sky Survey (DSS) images - throughout this
paper we will use the designations of \HI\ detections from D97: ce-70, ce-86, ce-102,
ce-137, ce-224, sw-89, sw-103, sw-146, sw-174, sw-194, sw-201 and 47-52. Three of
these (ce086, sw-103 and sw-194) will not be discussed in the present paper, for
reasons explained below.

At the positions of two of these \HI\ detections invisible on the DSS, sw-103 and sw-194, 
faint optical counterparts were detected in the deeper $B$, $V$ and $i$-band CCD images 
we obtained of all reported \HI\ clouds, except 47-52 (Duc et al. 2001; Iglesias-P\'aramo 
et al. 2002, hereafter IP02).  Although we could not obtain optical redshifts of  
sw-103 and sw-194 they are the only galaxies seen within the \HI\ clouds and their 
optical centre positions are only 4$''$ to 11$''$ (one fourth and half the VLA HPBW,
respectively) away
from that of the \HI\ cloud centres. We therefore regard them as tentative optical detections
of these two \HI\ clouds.

Given our non-confirmation at Arecibo of the reported \HI\ clouds, it should be noted
that we did clearly confirm the D97 VLA \HI\ line signal from, e.g., sw-103, which is as weak 
as the faintest reported VLA \HI\ cloud profiles, during the same observing run at Arecibo (IP02).

A third \HI\ cloud, ce-86, although it shows a strong line in 
the VLA survey and remained undetected on our CCD images, is clearly part of the extended 
\HI\ distribution of the peculiar galaxy IC 1182, and it therefore cannot be grouped 
together with the other clouds. Our Arecibo and optical observations of these three objects are
discussed in IP02. 

We failed to detect any optical emission from the remaining 8 \HI\ clouds covered by our 
images (47-52 was not observed) to a limiting V-band surface brightness of about 
27 mag arcsec$^{-2}$. For a contour 
plot of the D97 VLA \HI\ column density maps superimposed on the V-band images of IP02, 
showing all of the \HI\ clouds under consideration in the present Paper, except 47-52, 
see Figure 1.

The present paper concerns the \HI\ line observations we made at Arecibo, the only
other radiotelescope with the sensitivity required for checking
the reality of the tentative D97 \HI\ detections without optical counterparts.
The Arecibo observations are described in Section 2, and their results in Section 3.
In Section 4 the results are discussed and the conclusions presented.

\section{Observations and data reduction} 
We made our observations with the
refurbished 305 m Arecibo Gregorian radio telescope in May and June 2002. 
Data were taken with the L-Band Narrow receiver (circularly polarized at 1415 MHz),
using  nine-level sampling with two of the 2048 lag sub-correlators set to
polarization A and two to polarization B. All observations were taken in the
position-switching mode, with the off-source observation taken for the same length of
time and over the same portion of the Arecibo dish as the on-source observations. Each
on+off pair was followed by a 10 seconds on+off observation of a noise diode
calibration source. In principle, each scan consisted of a 5 minute on/off pair. The
total net integration time (on+off) was on average 90 minutes per source, the maximum
being 140 minutes for the faintest reported VLA source, sw-089, and the minimum 40 for
sw-174, where a nearby strong continuum source made sensitive observations impossible.
Each cloud was observed with each of the 4 sub-correlators centered at its redshifted
\HI\ line frequency. Two of the sub-correlators, one per polarization, were set to a
12.5 MHz bandpass, resulting in a velocity coverage of about 2500 km/s and a velocity
resolution of 1.3 km/s, while the other two were set to a 3.25 MHz bandpass. The
telescope's HPBW at 21 cm is \am{3}{4}$\times$\am{3}{6}. For the telescope's pointing 
positions the centre coordinates of the VLA \HI\ sources as given in D97 were used 
(see Table 1). For calibration purposes, a number of strong
continuum sources as well as spiral galaxies with strong \HI\ lines from the catalogue
of Lewis (1983) were observed throughout the run.

The observations were made at a mean frequency of 1368 MHz, in the 1350-1400 MHz
frequency band allocated on a co-primary basis to the Radio Astronomy Service, where
its protection from harmful interference is limited. Though care was taken to make the
renovated Arecibo telescope more robust against radio frequency interference (RFI),
and to coordinate its operation as well as possible with the frequency plan and
emission periods of local radar installations, RFI signals with strengths that hamper
the detection of faint \HI\ line signals were present during the first half of the
observations. After the main terrestrial RFI source was identified halfway through the
observing run, a blanker could be implemented that effectively removed these signals
from our spectra.

The data were reduced using IDL routines developed at Arecibo Observatory. The two
polarizations were averaged and corrections were applied for the variation in gain and
system temperature of the telescope as function of azimuth and zenith angle, using the
most recent calibration data available. A first-order baseline was then fitted to the
data, excluding those velocity ranges with \HI\ line emission or RFI. Once the
baselines were subtracted, the velocities were corrected to the heliocentric system,
using the optical convention. All data were boxcar smoothed to a velocity resolution
of 19.5 \kms\ for further analysis. If the average of all data on an object contained
a signal with a peak level exceeding 1 mJy, it was checked if that feature
corresponded to RFI signals that did not occur in all spectra - if so, the
contaminated spectra were not used for the final analysis. The rms noise levels of the
averaged spectra were determined in channels 300-1800, avoiding lines with a maximum
exceeding about 1 mJy peak line flux density.


\begin{figure*}[h]
\centering
\includegraphics[width=13.5cm]{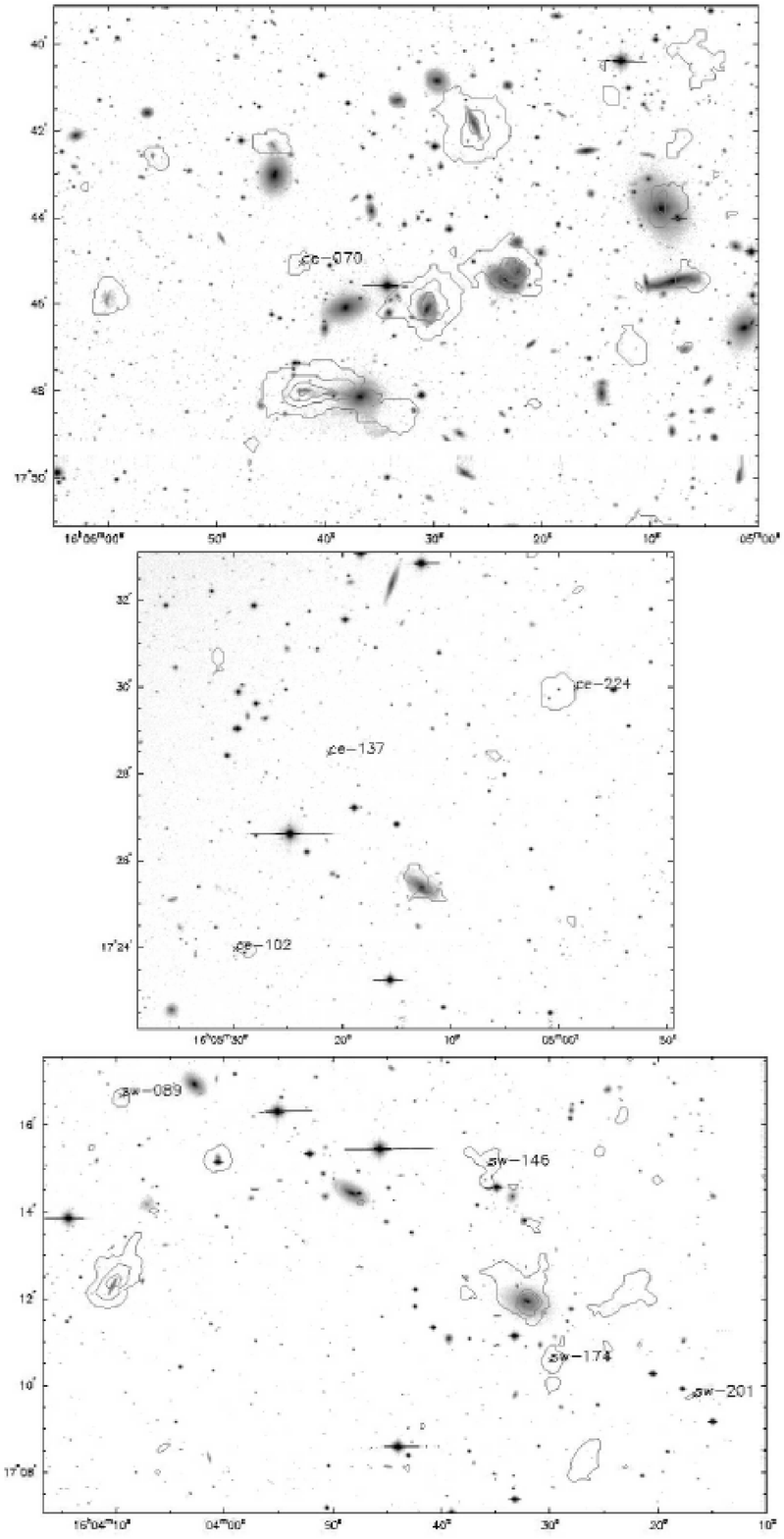}
\caption{\HI\ column density contours from the Dickey (1997) VLA survey superimposed
on our $V$-band CCD images (from Iglesias-P\'aramo et al. 2002). 
The \HI\ clouds without
optical counterparts in these images, reported as tentative detections in Dickey
(1997), have been indicated. }
 \end{figure*}

\section{Results}  
The Arecibo \HI\ line spectra of 8 tentative \HI\ clouds, smoothed to a resolution of
19.5 \kms\ are shown in Figure 2. The exception is sw-174 (at
16$^{h}$03$^{m}$30$^{s}$, 17$^{\circ}$10$'$34$''$ (J2000.0) - see Figure 1), for which
no sensitive \HI\ observations could be obtained due to the proximity, at \am{2}{2}
separation, of the elliptical galaxy NGC 6034 with its extended, 713 mJy continuum
source.

%
%
\begin{table*}
\bigskip
{\scriptsize
\begin{tabular}{llrrrlrrrrrrr}
\multicolumn{13}{l}{{\bf Table 1.} Reported tentative \HI\ clouds without optical
counterparts -- basic 21cm line data.}\\
\smallskip \\
\hline
\vspace{-2 mm} \\
 & \multicolumn{11}{c}{------------------------------------------------------ VLA data ------------------------------------------------------} & Are. \\
Name  & \,\,\,R.A.\,\,\,  &  \,\,\,Dec.\,\,\,  &  $V_{HI}$ & $W_{50}$ & $I_{H}$ & $I_{ext}$ & $I_{peak}$ &
  $I_{int}$ & $S_{max,H}$ & $S_{ave,ext}$ & rms & rms \\
 & \multicolumn{2}{c}{\scriptsize (2000.0)} & {\scriptsize [km/s]} & {\scriptsize [km/s] }& 
  \multicolumn{4}{c}{\scriptsize  ------------ [Jy \kms]  ------------}& {\scriptsize [mJy]} & {\scriptsize [mJy]} 
 & {\scriptsize [mJy/beam]} & {\scriptsize [mJy]} \\
\vspace{-2 mm} \\
\hline
\vspace{-1 mm} \\
ce-70  & 16 05 42.2 & 17 45 03 & 11654 & 396 & 0.25 & 0.53 & 0.19 & 1.22 & 1.0 & 1.35 & 0.17 &  0.29 \\ 
ce-102 & 16 05 29.8 & 17 23 57 & 11300 & 485 & 0.69 & 1.28 & 0.59 &  1.75 & 2.0 & 2.64 & 0.48 &  0.48 \\ 
ce-137 & 16 05 21.5 & 17 28 27 & 11625 & 351 & 0.36 & 0.92 & 0.37 & 0.34 & 1.5 & 2.63 & 0.30 &  0.39 \\ 
ce-224 & 16 04 58.5 & 17 29 58 & 11941: & 171: & 0.21: & 0.52: & 0.14: & 4.71: & 1.4 :& 5.85:  & 0.25 & 0.39 \\ 
sw-089 & 16 04 09.5 & 17 16 41 & 11374 & 528 & 0.29 & 0.41 & 0.22 & 0.32 & 0.8 & 0.78 & 0.15 &  0.25 \\ 
sw-146 & 16 03 35.6 & 17 15 05 & 10424 & 395 & 0.30 & 0.56 & 0.12 & O.17 &1.1 & 1.43 & 0.15 &  0.44 \\ 
sw-201 & 16 03 16.6 & 17 09 46 & 10711 & 484 & 0.44 & 0.71 & 0.17 & 0.11 & 1.3 & 1.47 & 0.18 &  0.31\\ 
47-52 & 16 03 16.2 & 16 14 21 & 10136 & 305 & 0.63 & 0.74 & 1.04 & 0.42 &2.5 & 2.43 & 0.58 & 0.38 \\ 
\vspace{-2 mm} \\
\hline
\end{tabular}
}
\normalsize
\end{table*}

Global \HI\ properties of the 8 observed clouds for which proper line spectra could be
obtained at Arecibo are listed in Table 1. Most data listed are taken from, or based
on, the D97 VLA observations. These data have a
synthesized beam size (HPBW) varying from 20$''$$\times$21$''$ to
26$''$$\times$29$''$, 
a velocity resolution of 88.4 \kms\ after Hanning smoothing
and an rms noise of about 0.13 mJy/beam per channel map at the field centre, which has been
multiplied by the primary beam attenuation factor to indicate the actual rms noise levels 
at the location of each reported \HI\ cloud in Table 1. 
With a pixel size of 6$''$$\times$6$''$, there are
about 24 pixels per synthesized beam. All radial velocities are heliocentric and
calculated according to the conventional optical definition,
$V=c$($\lambda$-$\lambda_0$)/$\lambda_0$.

Unless otherwise indicated, the following values were derived from the line profiles
shown in Figs. 9-12 in D97 and plotted here in Fig. 2, which correspond to the
$I_{H}$ integrated line flux: 
Listed in the 12 columns of Table 1 are: 
(1) the working designation of the cloud from D97, 
(2 \& 3) the Right Ascension and Declination of the cloud's centre position, 
(4) the central velocity of the VLA line profile, 
(5) the FWHM of the profile, $W_{50}$,
(6)-(9) four different measures of the integrated \HI\ line flux ($I_{H}$, $I_{ext}$, 
$I_{peak}$ and$I_{int}$) , see the description below,
(10) the maximum flux density in the profile corresponding to $I_{H}$, $S_{max,H}$, and
(11) the mean flux density, $S_{ave,ext}$, defined as $I_{ext}$/$W_{50}$, which can be 
larger than $S_{max,H}$ (see below).
Also listed, in column 12, is the rms noise level of the Arecibo spectra measured at a
velocity resolution of 19.5 \kms.


\begin{figure*}
\centering
\includegraphics[width=17cm]{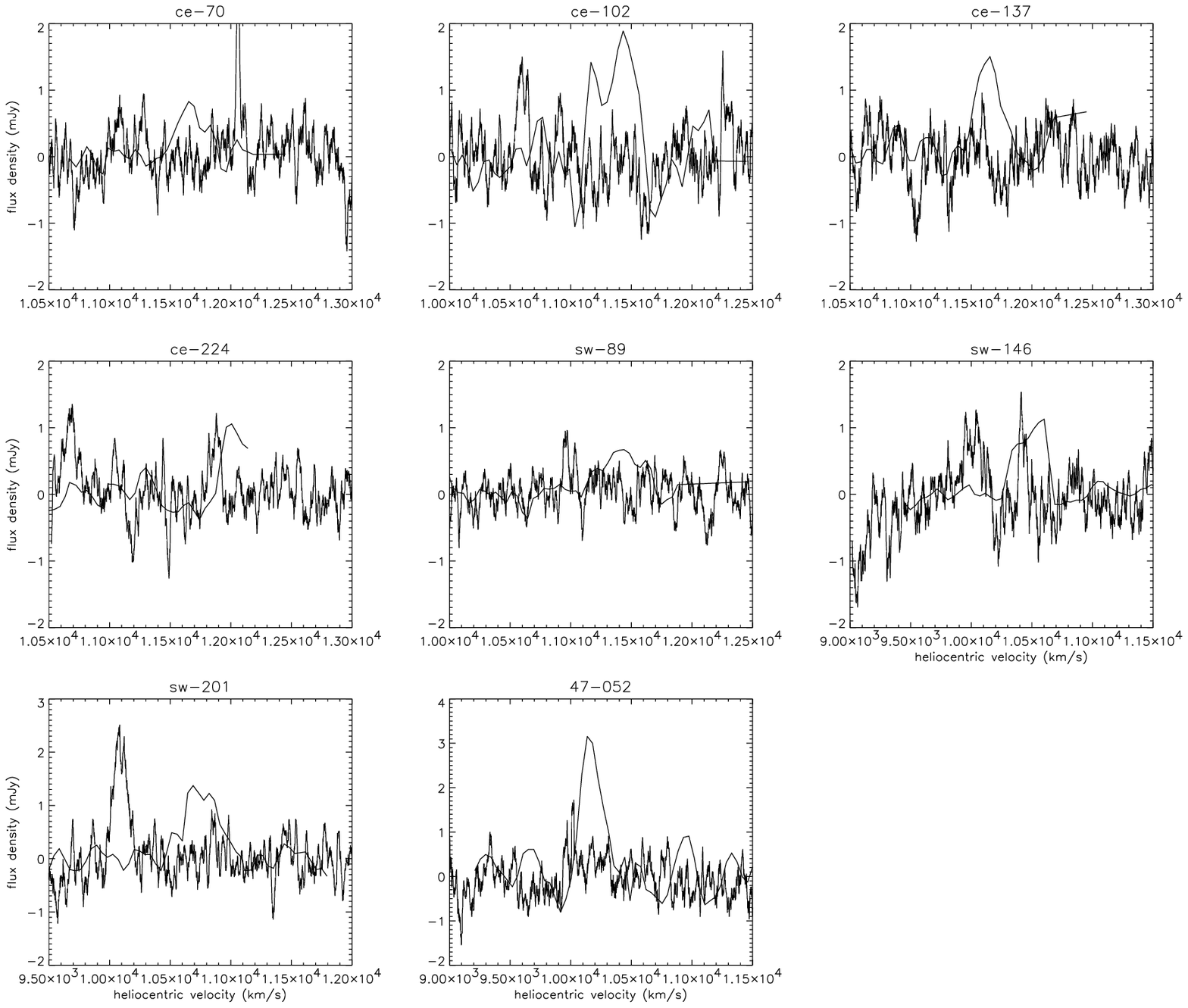}[h]
\caption{Arecibo 21 cm \HI\ line spectra of 8 reported (Dickey 1997) \HI\ clouds
without optical counterparts in the Hercules Cluster. The velocity resolution of the
data is 19.5 \kms. Shown also are the integrated VLA \HI\ line profiles from Dickey
(1997), with a velocity resolution of 88 \kms, corresponding to the $I_{H}$ integrated 
line flux measurements listed in Table 1.}
 \end{figure*}

Four different methods used for measuring integrated \HI\ line fluxes from
interferometric \HI\ line observations (listed in columns 8-11 of Table 1) lead to the
four different \HI\ masses listed in Table 2 of D97. We converted these \HI\ masses to
the integrated \HI\ line fluxes of Table 1 assuming a distance of 110.5~h$^{-1}$ Mpc
for the Hercules Cluster, following D97. The characteristics of these methods are as
follows: $I_{H}$ is the line flux measured by integrating the spectra over the group
of contiguous pixels above threshold, $I_{ext}$ is the line flux integrated over a
larger area, estimated by statistical tests to contain the total line emission, while
$I_{peak}$ and $I_{int}$ are obtained by fitting a two-dimensional Gaussian to the
velocity-integrated \HI\ column density map, where $I_{peak}$ corresponds to the line
flux within the central beam area and $I_{int}$ to the integrated flux of the
Gaussian. The latter is notoriously unstable, as can be seen by comparing the various
line fluxes listed in Table 1.

\subsection{Notes on individual \HIit\ clouds}  
Our observations of the \HI\ cloud ce-86 is discussed in IP02, as part of our Arecibo 
\HI\ and optical search for dwarf galaxies in the Hercules Cluster, since this cloud 
belongs to the extended \HI\ distribution connected to the peculiar galaxy IC 1182 
(see Figure 1). It seems to be the counterpart of  the tidal dwarf galaxy ce-61 on the opposite side 
of IC 1182, whose dynamical nature as a gravitationally bound system inside the tidal 
tail of the merger system has been confirmed recently through H$\alpha$
line Fabry-P\'{e}rot imaging (Duc \& Amram 2003; see also IP02 and Braine et al. 2001). 

A search was made for \HI\ signals that could possibly lead to confusion with the Arecibo 
spectrum of a target object, using the D97 VLA maps, within a 10$'$ radius around each pointing 
position. Recent measurements of the Arecibo antenna beam pattern, including sidelobes, were
used to estimate which percentage of the line flux of nearby objects would be detected.

{\bf ce-70: }\, The VLA detection is a bit tentative, as it is not very strong and there are only 14
pixels above the detection threshold (D97). 
No confusion from nearby galaxies is expected within the velocity range covered by the D97 \HI\ 
profile of this cloud. The two $\sim$0.8 mJy peaks seen in the 11,050-11,350 \kms\ 
range are probably due to a number of galaxies at about 3$'$-6$'$ distance, which all have redshifts
between 11,100-11,189 \kms: ce-48, 60, 95 and 109. 

{\bf ce-102: }\, The VLA detection is very tentative: though the line is not weak, there are only 13
pixels above the detection threshold and the emission is weakened by a factor 3.7 due to primary beam
attenuation (D97). No nearby galaxy is expected to cause confusion within the D97 \HI\ profile
of this cloud. There is no obvious candidate for the 1.3 mJy peak at $\sim$10,600 \kms\ in the Arecibo
data.

{\bf ce-137: }\, The VLA detection is tentative, as the line is weak, there are only 17 pixels above the
detection threshold and the emission is weakened by a factor 2.3 due to primary beam attenuation (D97).
The \HI\ source ce-155, at \am{1}{2} separation, could, in principle, be a source of confusion with the
profile of this \HI\ cloud, as its VLA line parameters are $V_{HI}$= 11,620 \kms, $W_{50}$= 551 \kms\ 
and $I_{ext}$= 1.0 Jy \kms\ (D97). ce-155 is classified as a tentative detection in D97, however, for the same
reasons as ce-137. There is no significant line signal detected at Arecibo within the range of
the VLA profile of this \HI\ cloud.

{\bf ce-224: }\, Rather tentative VLA detection, as the bandpass ends at 12,000 \kms, above which the
spectrum becomes so noisy that it is impossible to say what it is doing (D97).
The central VLA \HI\ velocity of $\sim$11,941 \kms\ is therefore uncertain, as are all other profile
parameters from D97, and it may be an underestimate - see Figure 2.
Only slightly overlapping with the VLA profile of the \HI\ cloud, the Arecibo spectrum 
shows a 1.2 mJy peak with $V_{HI}$= 11,849 \kms, $W_{50}$= 163 \kms\ and 
$I_{HI}$= 0.096 Jy \kms, which the Arecibo sidelobe pattern at the time of our observations shows 
may possibly be due to Sc spiral ce-294, at \am{7}{3} separation, 
with VLA line parameters $V_{HI}$= 11,875 \kms, $W_{50}$= 306 \kms\ and $I_{ext}$= 0.81 Jy \kms\ (D97).
There is no nearby galaxy in the VLA data that could have caused the $\sim$1.3 mJy peak around
10,700 \kms\ in the Arecibo data.

{\bf sw-089: }\, The VLA detection is extremely tentative, as the line is weak and broad and there are
only 5 pixels above the detection threshold (D97). There are no  nearby galaxies expected to cause 
confusion with the D97 VLA profile of this \HI\ cloud. The Arecibo data show two $\sim$0.7 mJy peaks
with $V_{HI}$= 10,995  \kms, $W_{50}$= 129 \kms\ and $I_{HI}$=  0.067 Jy \kms\, which may well be due to 
galaxy ce-103, at \am{2}{5} separation, for which we measured $V _{HI}$= 10,998 \kms, $W_{50}$= 155 \kms\ 
and $I_{HI}$= 0.17 Jy \kms\  at Arecibo (IP02).

{\bf sw-146: }\, Though a faint red image coincident with the \HI\ cloud was noted on the DSS in D97,
no galaxy was found at this location on our deeper CCD images (IP02). The VLA detection is somewhat
suspect as the baseline is somewhat degraded due to the presence of the strong (713 mJy) continuum
emission from nearby NGC 6034 (D97).
In the velocity range of the VLA profile, the Arecibo spectrum shows a 1.5 mJy peak with 
$V_{HI}$=  10,451 \kms, $W_{50}$=  176 \kms\ and $I_{HI}$= 0.042  Jy \kms. There are no nearby
galaxies in the VLA data that could have caused this peak, nor the two $\sim$1.2 mJy peaks around
10,000 \kms. We consider that the reported VLA cloud sw-146 has not been confirmed by the Arecibo
data. This Arecibo spectrum has a less smooth baseline than the others and the peaks might
be spurious and due to RFI.

{\bf sw-174: }\, The strong (713 mJy), extended continuum source in the nearby galaxy NGC 6034
made sensitive Arecibo line observations impossible and the VLA detection tentative.
The VLA profile is weak (0.7 mJy peak in the $I_{H}$ profile) and there are only 6 pixels 
above the detection threshold. The ``line'' could be a figment of imperfect bandpass calibration 
and continuum subtraction, as the continuum emission of NGC 6034 is sufficiently strong to increase 
the noise in the spectral baselines in its vicinity (D97). 

{\bf sw-201: }\, The VLA detection is very tentative, as there are only 6 pixels above the
detection threshold (D97).  There is no nearby galaxy that is expected to cause confusion
with the VLA profile. The Arecibo data show $\sim$2.3 mJy peaks
with $V_{HI}$=  10,088 \kms, $W_{50}$=  220 \kms\ and $I_{HI}$=  0.35 Jy \kms,
which appear due to nearby galaxy sw-222, at \am{2}{6} separation, for which we  measured 
$V _{HI}$= 10,079 \kms, $W_{50}$= 215 \kms\ and $I_{HI}$= 0.76 Jy \kms\  at Arecibo (IP02).

{\bf 47-52: }\, The VLA detection is extremely tentative, as there are only 9 pixels above the
detection threshold and the emission is weakened by a factor 4.4 due to primary beam attenuation
(D97). There are no nearby galaxies expected to cause confusion with the D97 VLA profile. 
Further smoothing in velocity of our spectrum only boosts the narrow peak at $\sim$10,000 \kms,
but not the signal in the velocity range of the VLA profile.

\section{Discussion and conclusions}  
Both single-dish and interferometric 21cm observations have their strong and weak points
in determining integrated line profiles of small, weak extragalactic \HI\ sources  like the 
tentatively detected \HI\ clouds we studied.

Although for a single-dish telescope like Arecibo observations result in only one spectrum
per pointing position and the derivation of integrated \HI\ profile parameters is straightforward,
these profiles depend on the instrument's beam pattern, which can lead to confusing
detections of nearby objects (see Section 3.1),
and single-dish data are more sensitive to RFI than interferometric data.

From interferometric data it is not straightforward to determine the integrated \HI\ line flux,
as pointed out in Section 3, hence the four different values listed for the VLA data in D97 
(see Table 1): of the flux determinations, $I_{H}$ and $I_{peak}$ can in principle be considered 
as lower limits, and $I_{ext}$ and $I_{int}$ as upper limits. Together, they 
give an indication of the uncertainties involved. A comparison of the various VLA fluxes with
our Arecibo fluxes for a sample of 20 small galaxies in the Hercules Cluster (from D97) shows 
(IP02) that the $I_{ext}$ fluxes are a reasonable estimate for the integrated
flux of weak \HI\ line signals detected over a small area: for the galaxies, on average, 
$I_{H}$ is 0.7$\pm$0.3 $I_{HI}$ (Arecibo) and  $I_{ext}$ is 1.0$\pm$0.5 $I_{HI}$. 
On average, the clouds' $I_{ext}$ fluxes are 1.9$\pm$0.5 times their $I_{H}$ VLA values.

Using the $I_{ext}$ line fluxes and the $W_{50}$ widths of the $I_H$ profiles of the clouds, 
in the absence of published widths of their $I_{ext}$ profiles, they have estimated average line
flux densities, $S_{ave,ext}$, varying between  0.8 and 5.8 mJy (mean 2.3 mJy). 
In comparison, the average rms noise level in our Arecibo data is  0.37 mJy 
at the used velocity resolution of 19.4 \kms, which is considerably smaller than
the 88.4 \kms\ of the VLA data.
We statistically expect to obtain signal-to-noise ratios 
of 6.2$\pm$3.5 for signals of the level of $S_{ave,ext}$.
Thus, for 6 out of the 8 observed sources we would expect to obtain detections at Arecibo
with a signal-to-noise ratio of at least 4.5 for a signal at the mean $S_{ave,ext}$ level 
of the \HI\ line, while the remaining two have an expected ratio of about 3.2 for such a signal.

The reported VLA \HI\ cloud signals were not confirmed at Arecibo, however, as shown by 
a comparison of the profiles (Figure 2). The non-confirmation would be even more evident
if the plotted VLA profiles would be scaled by a factor of 1.9 in flux density (the average 
$I_{ext}$/$I_{H}$ line flux ratio).

On the other hand, VLA \HI\ lines from a score of small galaxies in the Hercules Cluster, 
some of which are as weak as those of the \HI\ clouds discussed here, were confirmed by 
us during the same Arecibo observing period (IP02). 
The two faintest confirmed  detections are of ne-204 and sw-103,
which have, for example, VLA $S_{max,H}$ of 0.9 and 0.85 mJy and $S_{ave,ext}$ of 1.2 and 0.9
mJy, respectively. A comparison with Table 1 shows that these flux densities are among the lowest 
values for the observed \HI\ clouds. 
This also indicates that the confirmation of the reported clouds should have been feasible at Arecibo. 

Surprisingly, the reported \HI\ clouds have rather broad VLA profiles, with $W_{50}$=390$\pm$110 \kms\
on average - equivalent to the linewidth of a fairly massive, edge-on spiral galaxy. 
From the Arecibo data we would derive a 5$\sigma$ upper limit of 4 10$^8$$h^{-2}$ \Msun\ to the \HI\ 
mass of a cloud in the Hercules Cluster with an assumed profile width of 75 \kms.

\acknowledgements{
We would like to thank the staff of Arecibo Observatory for their help with the
observations and data reduction, especially P. Perrillat, and the referee, S.E.
Schneider, for his comments.
We have made use of the NASA/IPAC Extragalactic Database (NED), which is
operated by the Jet Propulsion Laboratory, California Institute of
Technology, under contract with the National Aeronautics and Space
Administration.
}

\end{document}